# Quantum Hall Effect of Weyl Fermions in Semiconducting *n*-type Tellurene

Gang Qiu[1,2], Chang Niu[1,2], Yixiu Wang[3], Mengwei Si[1,2], Zhuocheng Zhang[1,2], Wenzhuo Wu[3] and Peide D. Ye[1,2]

[1]School of Electrical and Computer Engineering, Purdue University, West Lafayette, Indiana 47907, USA

[2]Birck Nanotechnology Center, Purdue University, West Lafayette, Indiana 47907, USA

[3]School of Industrial Engineering, Purdue University, West Lafayette, Indiana 47907, USA

⋆e-mail: Correspondence and requests for materials should be addressed P. D. Y. (yep@purdue.edu)

**Dirac and Weyl nodal materials can host low-energy relativistic quasiparticles. Under strong magnetic fields, the topological properties of Dirac/Weyl materials can directly manifest through quantum Hall states. However, most Dirac/Weyl nodes generically exist in semimetals without exploitable bandgaps due to their accidental band-crossing origin. Here we report the first experimental observation of Weyl fermions in a semiconductor. Tellurene, the 2D form of tellurium, possesses chiral crystal structure which induces unconventional Weyl nodes with a hedgehog-like radial spin texture near the conduction band edge. We synthesize high-quality n-type tellurene by a hydrothermal method with subsequent dielectric doping and detect a topologically non-trivial $\pi$ Berry phase in quantum Hall sequences. Our work expands the spectrum of Weyl matter into semiconductors and offers a new platform to design novel quantum devices by marrying the advantages of topological materials to versatile semiconductors.**

The emergence of graphene[1] has triggered vast interest in studying the properties of relativistic particles in low energy spectra of topological materials. For example, under high magnetic fields, the unconventional sequence of graphene quantum Hall states at filling factor $\nu = 4(n + \frac{1}{2})$ (here $n = 0, 1, 2 ...$) suggests the presence of Dirac fermions with $\pi$ Berry phase[2,3], which manifests the topological nature of Dirac points. Since then, many other classes of topological materials with Dirac/Weyl nodal features in their band structures have been predicted and identified[4,5] with great potential for spintronics, optoelectronics, and quantum computing applications. However, these Dirac/Weyl points generated by the crossing of either two bands or two branches of spin polarized bands are generically limited to semimetals without exploitable bandgaps. In this work we introduce a new semiconductor system: tellurene (two-dimensional form of tellurium) with Weyl nodal features in the vicinity of conduction band minima. The combination of topological materials and semiconductors in 2D limit allows us to explore the Weyl physics in a more controllable manner and to design topological devices.

Trigonal tellurium (Te) is a narrow bandgap semiconductor with a unique one-dimensional helical atomic structure. The crystal consists of three-fold screw-symmetric atomic chains interconnected by van der Waals forces, as shown in Fig. 1a. The screw symmetry distinguishes two irreducible enantiomers with opposite chirality in real space, which falls into either P3$_2$21 or P3$_1$21 space group, depending on the handedness. One prominent feature of the band structure occurs at the bottom of the conduction band at H point (Fig. 1b and 1c), where the spin degeneracy is lifted to form band crossing due to strong spin-orbit coupling (SOC) and broken inversion symmetry. Despite the similarity to the energy dispersion of a conventional Rashba band, Te exhibits a hedgehog-like radial spin texture with positive (negative) monopole charge in momentum space arising from the topologically non-trivial Weyl node[6-10], which is fundamentally distinct from tangential spin texture Rashba bands[6] (as we shall further discuss later). The existence of Weyl points has been predicted recently[6-8,10], with topological properties in accordance with the framework of newly proposed Kramers-Weyl nodes in chiral crystals with strong SOC[11]. In chiral crystal systems like CoSi and $Ag_2Se$, angle-resolved photoemission spectroscopy[12,13] and magneto-transport[14] provide evidence for Kramers-Weyl fermions. Note that in Te since the H point is not a time reversal invariant momenta point, the degeneracy is not protected by Kramers theorem (which is the prerequisite of Kramers-Weyl nodes), but is instead guaranteed by its three-fold rotational symmetry. Nonetheless, despite the different mechanism leading to Weyl nodes, they both share similarity in band topology, spin textures and other topological properties.

To date, most magneto-transport measurements were performed on bulk Te samples[15-17], where the resolution of Shubnikov-de Haas (SdH) oscillation features was limited by the sample quality as well as weak two-dimensional (2D) Landau quantization. A recently developed hydrothermal growth method yields atomically flat 2D Te films with dangling-bond-free surfaces[18-20] (Fig. 1e inset), whose thickness ranges from a couple of atomic layers to tens of nanometers. These 2D Te films are coined as tellurene in analogy to other 2D elemental materials such as graphene, stanene and phosphorene. Much better developed SdH oscillations as well as quantum Hall states were observed in these p-type tellurene films[21], adding a new member into the scarce family of high-mobility 2D materials hosting the quantum Hall effect which includes graphene[2,3,22], black phosphorous[23-25], InSe[26], and some transition metal dichalcogenides[27-30]. However, due to the unintentional p-type doping in Te, the chemical potential is usually pinned near the valence band edge. Therefore so far most of the experiments have been limited to holes and little about the properties of the conduction band is revealed via transport measurements[19,21,31]. Here we successfully converted tellurene films into n-type using an atomic layer deposited (ALD) dielectric doping technique without degrading electron mobility[32,33], which grants us the access to the conduction band to explore more exotic physics of Weyl fermions through quantum Hall states. A schematic of a n-type doped device is illustrated in Fig. 1d. The as-synthesized tellurene films were dispersed onto a degenerately doped silicon substrate with a 90 nm $SiO_2$ insulating layer on top, followed by patterning and deposition of titanium/gold (Ti/Au) metal contacts. We chose low work function metal contacts (Ti) to lower the electron Schottky barrier height and accommodate n-type transport. A layer of alumina was subsequently deposited onto tellurene film by ALD grown at 200 ºC, converting the tellurene film from p-type to n-type, as confirmed by $I_d$-$V_g$ transfer curves of a tellurene field-effect transistor at both room temperature and cryogenic temperature (see Fig. 1e). Similar ALD doping method has also been reported on other material systems like black phosphorus[34-36] and silicon[37]. The doping mechanism is commonly attributed to the threshold voltage shift caused by positive fixed charges in low-temperature ALD-grown films[34] or the interface electric dipole field[37].

For magneto-transport measurements, six-terminal Hall bar devices were fabricated as shown in Fig. 2a (see Methods for more fabrication and measurement details). We investigated over 20 devices with typical film thickness ranging from 10 to 20 nm, and all of them exhibit similar and reproducible behaviors. Here all the data presented is from one 12-nm-thick high-quality device, unless otherwise specified (additional data set from other devices are available in Supplementary Note 1). The global back gate allows us to tune the 2D electron density from $2\times10^{12}$ to $1\times10^{13}$ cm$^{-2}$. Representative longitudinal ($R_{xx}$) and transverse ($R_{xy}$) magnetoresistance curves measured at $V_g$=10 V are plotted in Fig. 2b, with Hall density of $2.5\times10^{12}$ cm$^{-2}$ and Hall mobility of 6,000 cm$^2$/Vs. The onset of SdH oscillations is around 2 T, leading to an estimated quantum mobility of 5,000 cm$^2$/Vs, which is close to Hall mobility within a reasonable margin. At a B field of around 24 T and 32 T, $R_{xy}$ is fully quantized into an integer fraction of h/e$^2$ (corresponding to filling factor $\upsilon$=3 and 4), and $R_{xx}$ also drops to 0 -- a hallmark of the quantum Hall effect[38]. As shown in Fig. 2c, by fixing the B field at 42 T and sweeping the back gate voltage, all filling factors from 2-8 are resolved (although not all fully quantized due to Landau level broadening), suggesting all the degeneracies, including spin and valley, have been lifted. By mapping out $R_{xx}$ through the $V_g$-B parameter space, we can construct the Landau fan diagram in a colour map as in Fig. 2d. Since the conduction band edge is located at two inequivalent H (H') points in the first Brillouin zone accounting for two-fold valley degeneracy, it is conceivable that each Landau level consists of four degenerate energy states (2 for valley degeneracy $g_v$ and 2 for spin degeneracy $g_s$) like graphene, with cyclotron energy gap

$$E_C = \hbar\omega_c = \frac{\hbar eB}{m^*} \tag{1}$$

at filling factors of 4, 8, 12, etc. Here $\hbar$ is the reduced Planck constant, e is the electron charge, and $m^*$ is the effective mass. Following this argument, we should expect the energy gap to increase monotonically in the sequence of 4n as we approach lower filling factors, since the cyclotron energy $\frac{\hbar eB}{m^*}$ increases linearly with the magnetic field. Yet we notice from Fig. 2b that the gap at $\nu = 12$ is significantly larger than that at $\nu = 8$, suggesting the simple four-level single particle picture cannot explain these unconventional sequences.

To understand this anomaly, we first focus on SdH oscillation features in the relatively low B field regime. The blue curve of Fig. 3a shows the normalized longitudinal resistance $R_{xx}$ versus magnetic field (0 to 12 T) in another high mobility sample. Besides the predominant set of oscillations (marked by black arrows), a secondary set of weaker oscillations is also resolved. Due to the low oscillation frequency and limited observation window, it is difficult to disentangle two sets of oscillations by simply performing fast Fourier transform. Hence, we use the complete Lifshitz-Kosevich (L-K) formula to fit the SdH experimental data and distinguish two sets of oscillation features. The superposition of two sets of SdH oscillations in single electron picture can be described by:

$$\frac{\Delta R_{xx}}{R_{xx}} = \sum_{D=1,2} \sum_{r=1}^{N_r} \frac{r\lambda}{\sinh(r\lambda)} \exp\left(\frac{-r\pi m^*}{\tau_D eB}\right) \cos\left(r\left(2\pi \frac{B_{f,D}}{B} - \pi + \varphi_D\right)\right) \quad (2)$$

Here the first sum over D refers to two sets of oscillations, and higher order harmonics up to $N_r = 20$ were taken into account in our fitting. $\lambda = 2\pi^2 m^* k_B T/\hbar eB$ is the thermal damping term. We also parameterized the carrier lifetime $\tau_D$, oscillation frequency $B_{f,D}$ and phase offset $\varphi_D$ for both sets. The resistance minima are evenly spaced in $1/B$ plot, hence the period in $1/B$ can be translated into SdH oscillation frequency $B_f = \frac{1}{\Delta(1/B)}$. The best fitting results (red curve in Fig. 3a) yield a dominant set of oscillations with frequency $B_{f,1} = 38.9\ T$, carrier lifetime $\tau_1 = 0.28\ ps$ and phase offset $\varphi_1 = 0.97\ \pi$. The secondary set of oscillations were also well captured with frequency $B_{f,2} = 35.8\ T$, carrier lifetime $\tau_1 = 0.20\ ps$ and phase offset $\varphi_1 = 1.15\ \pi$. We conjecture that two sets of independent oscillations arise from the unique ALD doping scheme of the n-type Te films. The fixed positive charges in the ALD dielectrics will attract a layer of electrons on the top surface with relatively lower carrier mobility (and shorter lifetime) due to low interface quality and the back gate will induce another layer of charge on the bottom surface, as indicated in Fig. 3b. Following the same L-K formula, we can extend our simulation of SdH oscillations to the entire $V_g$-B parameter space, by assuming that the oscillation frequency for each set has a linear dependence on the back-gate bias, and meanwhile we set both phase shifts to $\pi$, which we shall further validate later. We intentionally exaggerated the carrier mobility in order to highlight the Landau levels. Our calculation (Fig. 3e) successfully reproduced the major features in the experiment (Fig. 3d). The oscillation frequency of the predominant (bottom layer) set increases from 17.9 T to 57.8 T as the back-gate sweeps from 10 V to 40 V; whereas the weaker set (top layer) increases from 21.7 T to 46.5 T. This is consistent with our dual-layer assumption as the frequency of the predominant set associated with the bottom layer is more sensitive to the back-gate modulation as illustrated in Fig. 3c, whereas the top layer with lower mobility is less sensitive since the gate-induce electric field is partially screened out by the bottom layer of electrons.

The most significant feature from the mapping in Fig. 3d and 3e is that the $V_g$-B parameter space can only be accurately reproduced when we fix the phase offset $\varphi_D$ at $\pi$, not 0. Furthermore, the purple and magenta dashed lines in Fig. 3d plotted from $B_n = B_f(V_g)/n$ (where $B_f(V_g)$ is the oscillation frequency under gate bias $V_g$ and $n$ is a positive integer) coincide with the bright strips which represent maxima in the oscillations. This instantly distinguishes Te from other trivial semiconductor systems. The Onsager-Lifshitz quantization rule dictates the oscillation amplitude to take the form:

$$\Delta R_{xx} \propto \cos\left(2\pi(B_F/B + 1/2) + \varphi\right) \qquad (3)$$

, and thus in a trivial semiconductor system where $\varphi = 0$, the $N$-th Landau level should be observed at field $B_n = B_f(V_g)/(n + \frac{1}{2})$. This suggests that the SdH oscillations in the Te system possess a phase offset $\varphi = \pi$, originating from the contribution of $\pi$ Berry phase near the Weyl node as we shall discuss next.

Evidence of Weyl fermions existing in bulk Te has been reported such as observations of negative magnetoresistance[39] and the kinetic magnetoelectric[10] effect. Here we present much more convincing evidence -- $\pi$ Berry phase in quantum oscillations. As a common practice, the phase offset is read off from the intercept of the Landau fan diagram. To construct the Landau fan diagram correctly, the resistivity $\rho_{xx}$ and $\rho_{xy}$ was first properly converted into conductivity $\sigma_{xx}$ and $\sigma_{xy}$ with an anisotropic Hall tensor (see Supplementary Note 2) to avoid phase shift[40] since the fundamental quantization occurs in conductivity and not resistivity[41,42]. The minima in conductivity $\sigma_{xx}$ are assigned with integer index n and the maxima are identified with half integers, and then plotted against 1/B. As shown in Fig. 4a and 4b, the extrapolated the linear fitting curves intercept the y-axis at $1/2$ throughout the entire gate range, which was also observed in many other Dirac or Weyl topological materials with non-trivial $\pi$ Berry phase[2,3,43-45]. We shall point out that it can be risky to directly associate a phenomenological phase offset in oscillations with Berry phase of Dirac/Weyl fermions in some scenarios [46,47] due to orbital moments and/or Zeeman effects. However, in a spin-orbit coupled system with time reversal symmetry but not spatial inversion symmetry like Te, the phase contributions from orbital moment and Zeeman effect cancel out due to symmetry. Therefore, it is safe to take the $\pi$ phase offset as a smoking gun for Weyl fermions in this case[48]. We can rule out another trivial cause of $\pi$ phase offset due to large Zeeman splitting (like in the case of $WSe_2$[49]) by absence of coincident effect when rotating the sample (see Supplementary Note 3). Furthermore, to assertively eliminate any potential offset from imperfection of the device, we measured another 8 devices and all of them unanimously show near $1/2$ intercept (see Supplementary Note 4).

The origin of the Berry phase is rooted in the hedgehog-like spin texture[6,9] near the chirality-induced Weyl nodes. Weak anti-localization effect is observed in the near-zero magnetic field regime (see Supplementary Note 5), indicating strong SOC in Te[50]. The strong SOC gives rise to a spin-polarized band crossing at the H and H' points (as shown in Fig. 4c), which is protected by the three-fold rotational symmetry. This point at H (and H') is classified as a Weyl node[6-9] that can be viewed as a Berry curvature monopole in the momentum space[8]. The spin textures are either pointing at or away from the Weyl nodes as illustrated by red arrows in Fig. 4c, which is in sharp contrast to a conventional Rashba SOC band whose energy dispersion may be similar, but whose spin texture is tangential (Fig. 4d). Under magnetic fields, when an electron completes a cyclotron motion in the real space its momentum changes by $2\pi$, corresponding to a closed orbit around the Fermi surface in momentum space. Along this path, the spin of the electron also rotates by $2\pi$, which picks up a $\pi$ Berry phase, since electrons are spin-1/2 particles. The hedgehog-like spin texture near Te Weyl nodes resembles that near the Dirac points of graphene which also give rise to a topologically non-trivial $\pi$ Berry phase, except that in graphene the Berry phase is induced by a radial pseudospin (valley isospin) texture [51,52] rather than the real spin as in Te.

We shall note that unlike normal band-inversion-induced Weyl semimetals which usually lead to linear band dispersion in a broader energy window, chirality-induced Weyl nodes can potentially exist near the band edge of a semiconductor system. These semiconductor systems with chirality-induced Weyl nodes grant us extended tunability to probe the topological properties of Weyl fermions by taking advantage of the versatility of semiconductors.

A sharp contrast between Te and other Weyl semimetals manifests in the effective mass measurement. In a nutshell, the Weyl nodes only guarantee the topological properties, but the band dispersion can still be arbitrary, depending the energy scale of interest. For example, Band-inversion-induced Dirac nodes in graphene are accompanied by linear band dispersion for the entire gate-accessible range near the Dirac points, since the first order of Hamiltonian can be accurately described by Dirac equation in the energy window near 1 eV. This gives rise to strongly gate-dependent cyclotron mass of relativistic Dirac fermions: $m_c = \frac{E_F}{v_F^2} = \sqrt{\pi \hbar^2 n_s / v_F^2}$, as confirmed by temperature-dependent SdH oscillation amplitudes[2,3]. Here we also extract cyclotron mass by fitting the temperature-dependent SdH amplitude with the thermal damping term in L-K formula (Fig. 5a):

$$\Delta R_{xx} \propto \frac{2\pi^2 m^* k_B T / \hbar e B}{\sinh(2\pi^2 m^* k_B T / \hbar e B)} \qquad (4)$$

The cyclotron mass is extracted to be 0.097±0.008 $m_0$ (where $m_0$ is the free electron mass) regardless of the gate bias (Fig. 5b), which is consistent with previously reported value of bulk Te[53,54]. This is because chirality-induced Weyl nodes reside only several meV above the conduction band minima and the semiconductor properties (band mass) is preserved. Meanwhile the topological properties (Berry phase) carry over to a much broader energy window of at least 50 meV that the gate can access.

In conclusion, using ALD dielectric doping on high-quality tellurene films, for the first time we observed well-developed quantum Hall effect of 2D electron gas in tellurene and accessed the electronic structure of its conduction band. A wide quantum well model with two layers of electrons is proposed to explain the anomaly in SdH oscillations and quantum Hall sequences. Spin and valley isospin degeneracies are fully lifted under 45 T external magnetic field, leading to fully polarized quantum Hall ferromagnetic states. Topologically non-trivial $\pi$ Berry phase is unambiguously detected and served as direct evidence of Weyl fermions in the vicinity of chirality-induced Weyl nodes with predicted hedgehog-like spin textures. N-type Te is an ideal material to explore the topological properties of relativistic quasiparticles in a chiral semiconductor system with tunable Fermi surface and exploitable band gap.

1. Geim, A. K. & Novoselov, K. S. The rise of graphene. *Nat. Mater.* **6**, 183–191 (2007).

2. Novoselov, K. S. *et al.* Two-dimensional gas of massless Dirac fermions in graphene. *Nature* **438**, 197–200 (2005).

3. Zhang, Y. B., Tan, Y. W., Stormer, H. L. & Kim, P. Experimental observation of the quantum Hall effect and Berry's phase in graphene. *Nature* **438**, 201–204 (2005).

4. Armitage, N.P., Mele, E.J. and Vishwanath, A. Weyl and Dirac semimetals in three-dimensional solids. *Rev. Mod. Phys.* **90**, 015001 (2018).

5. Hasan, M.Z., Xu, S.Y., Belopolski, I. and Huang, S.M. Discovery of Weyl fermion semimetals and topological Fermi arc states. *Annu. Rev. Condens. Matter Phys.* **8**, 289-309 (2017).

6. Hirayama, M., Okugawa, R., Ishibashi, S., Murakami, S. & Miyake, T. Weyl node and spin texture in trigonal tellurium and selenium. *Phys. Rev. Lett.* **114**, 206401 (2015).

7. Nakayama, K. *et al.* Band splitting and Weyl nodes in trigonal tellurium studied by angle-resolved photoemission spectroscopy and density functional theory. *Phys. Rev. B* **95**, 125204 (2017).

8. Tsirkin, S. S., Puente, P. A. & Souza, I. Gyrotropic effects in trigonal tellurium studied from first principles. *Phys. Rev. B* **97**, 035158 (2018).

9. Agapito, L. A., Kioussis, N., Goddard, W. A. & Ong, N. P. Novel family of chiral-based topological insulators: elemental tellurium under strain. *Phys. Rev. Lett.* **110**, 176401 (2013).

10. Şahin, C., Rou, J., Ma, J. & Pesin, D. A. Pancharatnam-Berry phase and kinetic magnetoelectric effect in trigonal tellurium. *Phys. Rev. B* **97**, 205206 (2018).

11. Chang, G. *et al.* Topological quantum properties of chiral crystals. *Nat. Mater.* **17**, 978 (2018).

12. Rao, Z. *et al.* Observation of unconventional chiral fermions with long Fermi arcs in CoSi. *Nature* **567**, 496–499 (2019).

13. Sanchez, D. S. *et al.* Topological chiral crystals with helicoid-arc quantum states. *Nature* **567**, 500–505 (2019).

14. Zhang, C. L. *et al.* Ultraquantum magnetoresistance in the Kramers-Weyl semimetal candidate β-$Ag_2Se$. *Phys. Rev. B* **96**, 1–10 (2017).

15. Klitzing, K. von & G.Landwehr. Surface quantum states in tellurium. *Solid State Commun.* **9**, 2201–2205 (1971).

16. Silbermann, R. & Landwehr, G. Surface quantum oscillations in accumulation and inversion layers on tellurium. *Solid State Commun.* **16**, 6–9 (1975).

17. Klitzing, K. von. Magnetophonon oscillations in tellurium under hot carrier conditions. *Solid State Commun.* **15**, 1721–1725 (1974).

18. Wang, Y. *et al.* Field-effect transistors made from solution-grown two-dimensional tellurene. *Nat. Electron.* **1**, 228–236 (2018).

19. Du, Y. *et al.* One-dimensional van der Waals material tellurium: Raman spectroscopy under strain and magneto-transport. *Nano Lett.* **17**, 3965−3973 (2017).

20. Wu, W., Qiu, G., Wang, Y., Wang, R. & Ye, P. Tellurene: its physical properties,

scalable nanomanufacturing, and device applications. *Chem. Soc. Rev.* **47**, 7203–7212 (2018).

21. Qiu, G. *et al.* Quantum transport and band structure evolution under high magnetic field in few-layer tellurene. *Nano Lett.* **18**, 5760–5767 (2018).

22. Gusynin, V. P. & Sharapov, S. G. Unconventional integer quantum Hall effect in graphene. *Phys. Rev. Lett.* **95**, 146801 (2005).

23. Li, L. *et al.* Quantum oscillations in a two-dimensional electron gas in black phosphorus thin films. *Nat. Nanotechnol.* **10**, 608–613 (2015).

24. Li, L. *et al.* Quantum Hall effect in black phosphorus two-dimensional electron system. *Nat. Nanotechnol.* **11**, 593–597 (2016).

25. Yang, J. *et al.* Integer and fractional quantum Hall effect in ultra-high quality few-layer black phosphorus transistors. *Nano Lett.* **18**, 229–234 (2018).

26. Bandurin, D. A. *et al.* High electron mobility, quantum Hall effect and anomalous optical response in atomically thin InSe. *Nat. Nanotechnol.* **12**, 223–227 (2017).

27. Fallahazad, B. *et al.* Shubnikov-de Haas oscillations of high-mobility holes in monolayer and bilayer $WSe_2$: Landau level degeneracy, effective mass, and negative compressibility. *Phys. Rev. Lett.* **116**, 1–5 (2016).

28. Movva, H. C. P. *et al.* Density-dependent quantum Hall states and Zeeman splitting in monolayer and bilayer $WSe_2$. *Phys. Rev. Lett.* **118**, 247701 (2017).

29. Wu, Z. *et al.* Even-odd layer-dependent magnetotransport of high-mobility Q-valley electrons in transition metal disulfides. *Nat. Commun.* **7**, 1–8 (2016).

30. Pisoni, R. *et al.* Interactions and magnetotransport through spin-valley coupled Landau levels in monolayer MoS$_2$. *Phys. Rev. Lett.* **121**, 247701 (2018).

31. Ren, X. *et al.* Gate-tuned insulator-metal transition in electrolyte-gated transistors based on tellurene. *Nano Lett.* **19**, 4738–4744 (2019).

32. Qiu, G. *et al.* High-performance few-layer tellurium CMOS devices enabled by atomic layer deposited dielectric doping technique. in *2018 76th Device Research Conference (DRC)* 1–2 (IEEE, 2018).

33. Berweger, S. *et al.* Imaging carrier inhomogeneities in ambipolar tellurene field effect transistors. *Nano Lett.* **19**, 1289–1294 (2019).

34. Liu, H., Neal, A. T., Si, M., Du, Y. & Ye, P. D. The effect of dielectric capping on few-layer phosphorene transistors: Tuning the schottky barrier heights. *IEEE Electron Device Lett.* **35**, 795–797 (2014).

35. Perello, D. J., Chae, S. H., Song, S. & Lee, Y. H. High-performance n-type black phosphorus transistors with type control via thickness and contact-metal engineering. *Nat. Commun.* **6**, 7809 (2015).

36. Wang, C. H. *et al.* Unipolar n-type black phosphorus transistors with low work function contacts. *Nano Lett.* **18**, 2822–2827 (2018).

37. Coss, B. E. *et al.* Near band edge Schottky barrier height modulation using high-κ dielectric dipole tuning mechanism. *Appl. Phys. Lett.* **95**, 222105 (2009).

38. Klitzing, K. V., Dorda, G. & Pepper, M. New method for high-accuracy determination of the fine-structure constant based on quantized Hall resistance. *Phys. Rev. Lett.* **45**, 494–497 (1980).

39. Zhang, N. *et al.* Evidence for Weyl fermions in the elemental semiconductor tellurium. *arXiv Prepr. arXiv1906.06071* (2019).

40. Ren, Z., Taskin, A. A., Sasaki, S., Segawa, K. & Ando, Y. Large bulk resistivity and surface quantum oscillations in the topological insulator $Bi_2Te_2Se$. *Phys. Rev. B* **82**, 241306 (2010).

41. Ando, Y. Topological insulator materials. *J. Phys. Soc. Japan* **82**, 1–32 (2013).

42. Xiong, J. *et al.* Quantum oscillations in a topological insulator $Bi_2Te_2Se$ with large bulk resistivity (6Ω·cm). *Phys. E Low-dimensional Syst. Nanostructures* **44**, 917–920 (2012).

43. Yu, W. *et al.* Quantum oscillations at integer and fractional Landau level indices in single-crystalline $ZrTe_5$. *Sci. Rep.* **6**, 35357 (2016).

44. Hu, J. *et al.* π Berry phase and Zeeman splitting of Weyl semimetal TaP. *Sci. Rep.* **6**, 18674 (2016).

45. Zhao, Y. *et al.* Anisotropic Fermi surface and quantum limit transport in high mobility three-dimensional Dirac semimetal $Cd_3As_2$. *Phys. Rev. X* **5**, 031037 (2015).

46. Roth, L. Semiclassical theory of magnetic energy levels and magnetic susceptibility of Bloch electrons. *Phys. Rev.* **145**, 434 (1966).

47. Dhillon, J.S., Shoenberg, D. The de Haas-van Alphen Effect III. Experiments at Fields up to 32 kG, *Phil. Trans. R. Soc. A* **248**, 1 (1955).

48. Alexandradinata, A., Wang, C., Duan, W. & Glazman, L. Revealing the topology of Fermi-surface wave functions from magnetic quantum oscillations. *Phys. Rev. X* **8**, 11027 (2018).

49. Xu, S. *et al.* Odd-integer quantum Hall states and giant spin susceptibility in p-type few-layer WSe$_2$. *Phys. Rev. Lett.* **118**, 067702 (2017).

50. Niu, C. *et al.* Gate-tunable Strong Spin-orbit Interaction in Two-dimensional Tellurium Probed by Weak-antilocalization. *arXiv Prepr. arXiv1909.06659* (2019).

51. Rotenberg, E. Topological insulators: The dirt on topology. *Nat. Phys.* **7**, 8–10 (2011).

52. Mallet, P. *et al.* Role of pseudospin in quasiparticle interferences in epitaxial graphene probed by high-resolution scanning tunneling microscopy. *Phys. Rev. B - Condens. Matter Mater. Phys.* **86**, 1–14 (2012).

53. Shinno, H., Yoshizaki, R., Tanaka, S., Doi, T. & Kamimura, H. Conduction band structure of tellurium. *J. Phys. Soc. Japan* **35**, 525–533 (1973).

54. Liu, Y., Wu, W. & Goddard, W. A. Tellurium: fast electrical and atomic transport along the weak interaction direction. *J. Am. Chem. Soc.* **140**, 550–553 (2018).

**Acknowledgements** P. D. Y. was supported by NSF/AFOSR 2DARE Program, ARO, and SRC. W. W. acknowledges the College of Engineering and School of Industrial Engineering at Purdue University for the startup support. W. W. was partially supported by a grant from the Oak Ridge Associated Universities (ORAU) Junior Faculty Enhancement Award Program. A portion of this work was performed at the National High Magnetic Field Laboratory, which is supported by National Science Foundation Cooperative Agreement No. DMR-1644779 and the State of Florida. G. Q and C. N acknowledge the technical support from NHMFL staffs J. Jaroszynski, A. Suslov and W. Coniglio. The authors want to give special thanks to K. von Klitzing, T. Ando, W. Pan, K. Chang, F. Zhang, C. Liu, K. Cho, Y. Nie and J. Hwang for the insightful discussion on electronic structures of Te. The authors also acknowledge A. R. Charnas for editorial assistance.

**Author Contributions** P. D. Y. conceived and supervised the project. P. D. Y. and G. Q. designed the experiments. Y. W. synthesized the material under the supervision of W. W. G. Q. and C. N. fabricated the devices. G. Q., C. N. and Z. Z. performed the magneto-transport measurements. G. Q., C. N., M. S., and Z. Z. analyzed the data. P. D. Y. and G. Q. wrote the manuscript with input and comments from all the authors.

**Author Information**

*School of Electrical and Computer Engineering, Purdue University, West Lafayette, Indiana, USA*

Gang Qiu, Chang Niu, Mengwei Si, Zhuocheng Zhang & Peide D. Ye

*Birck Nanotechnology Center, Purdue University, West Lafayette, Indiana, USA*

Gang Qiu, Chang Niu, Mengwei Si, Zhuocheng Zhang & Peide D. Ye

*School of Industrial Engineering, Purdue University, West Lafayette, Indiana, USA*

Yixiu Wang & Wenzhuo Wu

**Competing financial interests** The authors declare no competing financial interests.

**Figure 1| Crystal structure of Te and n-type ALD doped tellurene devices. a**, Crystal structure of Te with helical atomic chains. **b**, First Brillouin zone of Te. The conduction band minima are located at the corner points H and H'. **c**, Zoomed-in energy dispersion of lowest spin-split bands along K-H-A line. A Weyl node is formed at H point, as highlighted by the red dashed circle. **d**, schematic view of an n-type Te device with a global back gate and ALD doping layer on top. **e**, $I_d$-$V_g$ transfer curves of a typical ALD doped n-type device measured at room temperature (red) and cryostat temperature (blue). Inset: An optical image of an as-synthesized tellurene film obtained from hydrothermal growth method. The helical chains (z-axis) are aligned along the longer edge of the film, as indicated by the red arrow. All the Hall bar devices are also fabricated along z-axis.

**Figure 2| Quantum Hall effect in tellurene two-dimensional electron gas. a**, An optical image of a six-terminal Hall bar device. **b**, Longitudinal ($R_{xx}$, in blue) and transverse ($R_{xy}$, in red) resistance as a function of magnetic field. **c**, $\sigma_{xx}$ and $\sigma_{xy}$ a function of back gate voltage. The gate oxide is 90 nm $SiO_2$. **d**, Colour mapping of $R_{xx}$ by sweeping both back gate voltage and magnetic field. The data from 0-12T and 12-45T are measured from the same device in a superconducting magnetic system at 30 mK and a hybrid magnetic system at 300 mK, respectively.

**Figure 3| Shubnikov-de Haas oscillation analysis of two layers of electrons in Te wide quantum well. a**, Representative SdH oscillation features measured at 30 mK and $V_{bg}$=26 V (top) and fitting results from complete L-K formula with the superposition of two sets of independent oscillations. The black arrows mark the predominant set (bottom layer) and

the grey arrows represents the weaker set (top layer). **b**, Schematics of a back-gate device and carrier distribution on top and bottom of the Te film. **c**, Potential profile and wave function distribution of two layers of charge with increasing back gate (from red to blue). **d,** experimental and **e,** simulation of $R_{xx}$ mapping in $B$-$V_{bg}$ parameter space. The purple and light magenta dashed lines and labels on the side indicate Landau level index of back and top layers of electrons. The dashed lines are plotted by integer fraction of $B_f$.

**Figure 4| Weyl nodes and Berry phase near Te conduction band edge. a**, Landau fan diagram under different gate bias. The scattered symbols are read off from the value $1/B$ of minima in $\sigma_{xx}$ and plotted against the Landau level index. Straight lines are linear fitting under each gate bias. Inset: Magnified view of gray dashed box in **a** with linear fittings extrapolated to the y-axis. **b**, Intercept of linear fittings versus gate voltage. Error bars represent s.d. from linear fitting. The reference line at y=0.5 corresponds to π Berry phase. The energy dispersion and spin texture (red arrows) in **c**, Chirality-induced Weyl node at H point and **d,** a conventional Rashba spin-orbit coupling band.

**Figure 5| Temperature-dependent SdH oscillations and effective mass of Weyl fermions. a**, SdH oscillation amplitudes (subtracting a smooth background) under various temperature from 0.5 K to 18 K. Inset: $\Delta R_{xx}$ versus temperature fitted by thermal damping term of L-K formula. **b**, The effective mass is extracted to be about 0.097±0.008 $m_0$, and is independent of gate biases and magnetic fields. Error bars represents standard error from L-K formula fitting.

**Methods**

**Synthesis of 2D tellurium crystals**

In a typical procedure[18], analytical grade $Na_2TeO_3$ (0.00045 mol) and a certain amount of poly(-vinyl pyrrolidone) is resolved into deionized water (33 ml) at room temperature under magnetic stirring to form a homogeneous solution. The resulting solution is transferred into a Teflon-lined stainless-steel autoclave, which is then filled with an aqueous ammonia solution (25%, w/w%) and hydrazine hydrate (80 %, w/w%). The autoclave is sealed and maintained at the reaction temperature for a designed time. The autoclave is then allowed to cool to room temperature naturally. Silver-gray, solid tellurium products are precipitated by centrifugation at 5000 rpm for 5 minutes and rinsed 3 times with distilled water (to remove any ions remaining in the final product).

**Langmuir-Blodgett (LB) transfer of 2D Te films**

The hydrophilic 2D Te nanoflake monolayers can be transferred to silicon substrates by the Langmuir-Blodgett (LB) technique[55]. The washed nanoflakes are suspended in a mixing solvent of N,N-dimethylformamide (DMF) and $CHCl_3$ (e.g., in a ratio of 1.3:1). Then, the mixture is dispersed into deionized water. After the evaporation of the solvent, 2D Te flakes will be left floating on the surface of water. Then we can scoop the 2D Te films onto substrates.

**Device fabrication**

The as-grown 2D Te films were dispersed onto heavily doped Si substrates with a 90 nm $SiO_2$ insulating layer, followed by DI water rinse and standard solvent cleaning process (acetone, methanol and isopropanol). Hall-bar devices were patterned with electron beam lithography (EBL) and 30/90 nm Ti/Au metal contacts were deposited by electron beam evaporation at a pressure below $2\times10^{-6}$ Torr. To eliminate geometric non-idealities, the device was then trimmed into standard Hall bar shape with better symmetry using $BCl_3$/Ar dry etching. 20 nm of $Al_2O_3$ was deposited onto the Te films at 200 °C with atomic layer deposition to achieve n-type doping. For double-gated devices, another Ni/Au top metal gate was patterned and deposited with EBL and electron beam evaporation subsequently to cover the entire channel region.

**Magneto-transport measurements**

Low magnetic field transport measurements were performed using a Triton 300 (Oxford Instruments) dilution fridge system with 12 Tesla superconducting coils. The high magnetic field data was acquired in a 31 Tesla resistive magnet system (Cell 9) and a 45 Tesla hybrid magnet system (Cell 15) at the National High Magnetic Field Lab (NHMFL) in Tallahassee, FL. Electrical data was recorded using standard low-frequency AC measurement techniques using SR830 lock-in amplifiers. The transfer curves of two-terminal FETs were measured with a Cascade probe station and Keysight B1500A semiconductor analyzer at room temperature and current mode of a lock-in amplifier at 30 mK.

55. Zasadzinski, J. A., Viswanathan, R., Madsen, L., Garnaes, J. & Schwartz, D. K. Langmuir-Blodgett films. *Science* **263**, 1726–1733 (1994).

**Figure 1**

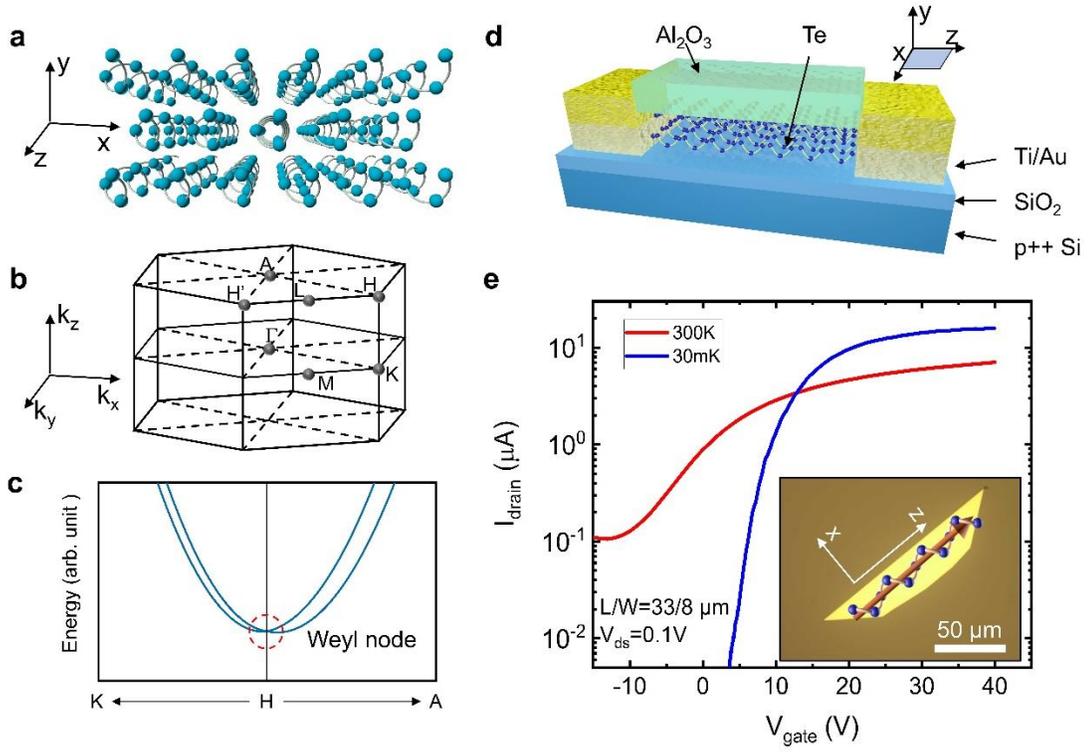

**Figure 2**

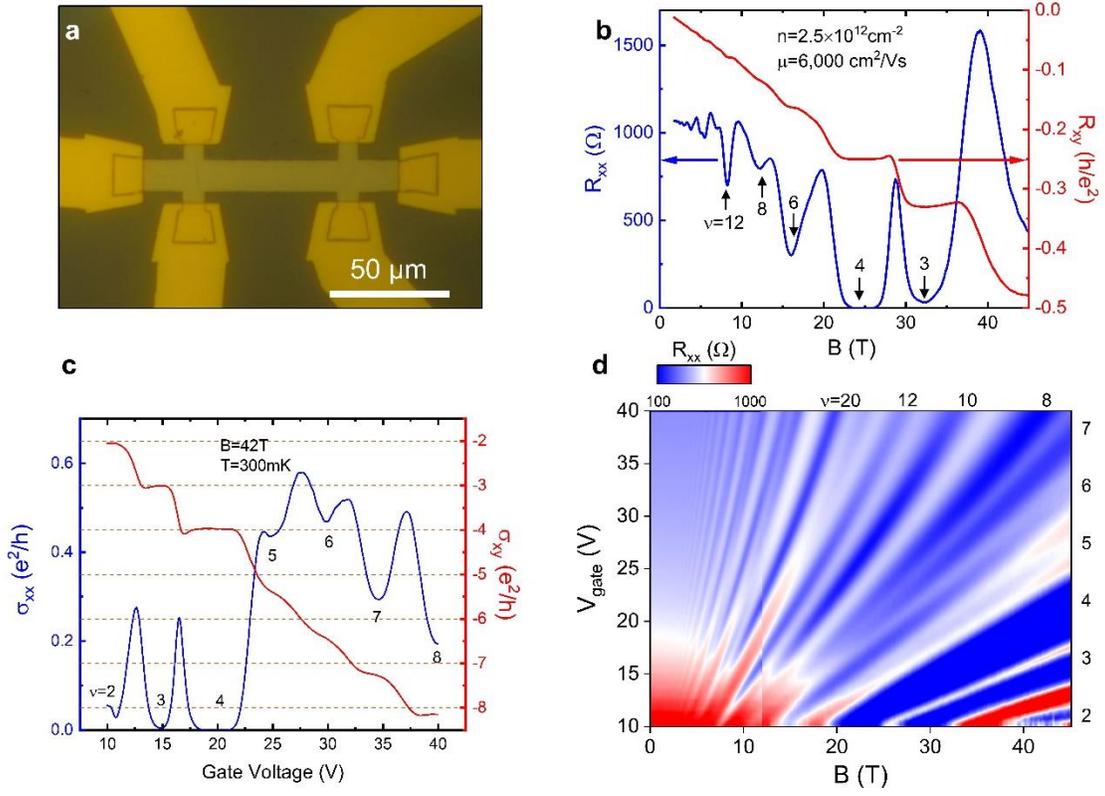

**Figure 3**

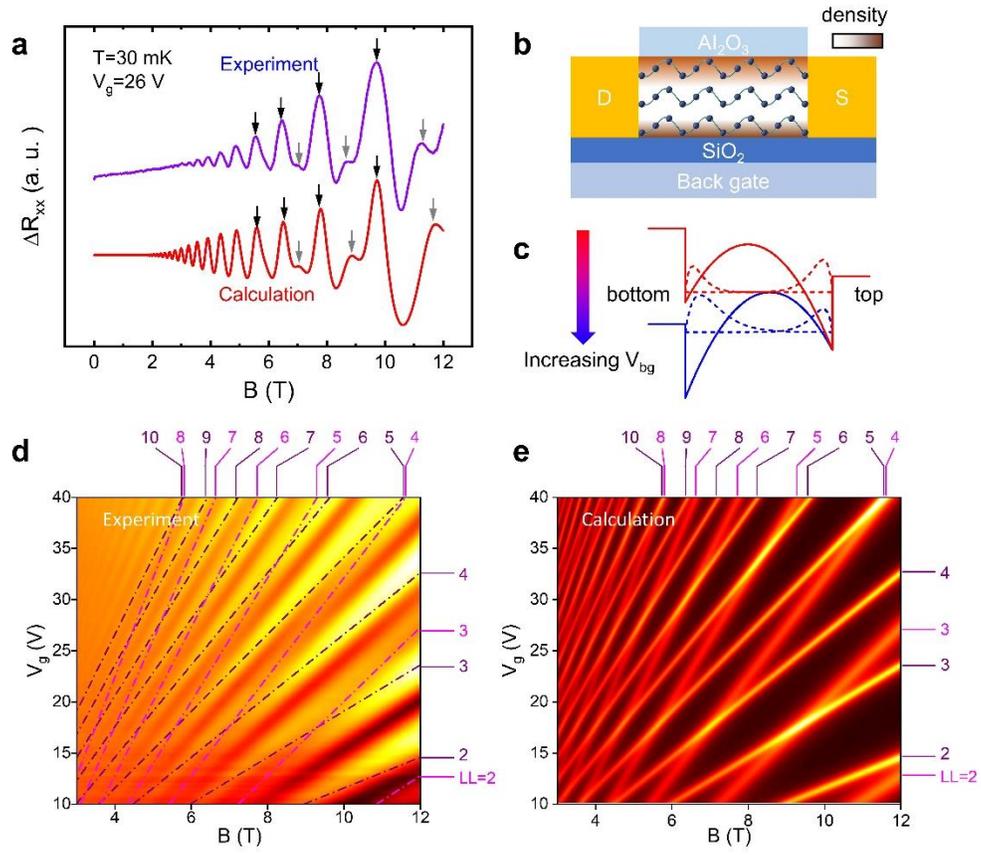

**Figure 4**

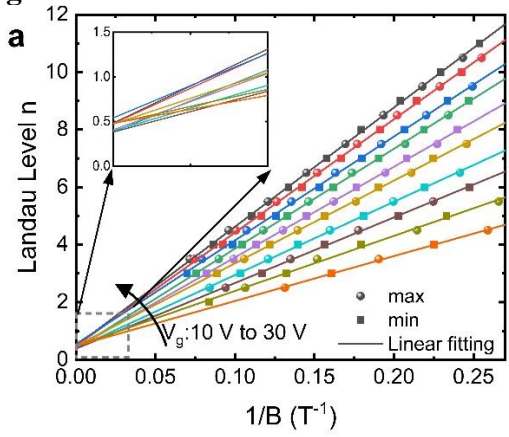
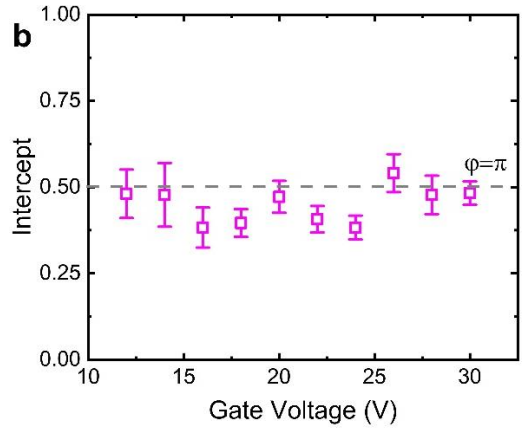
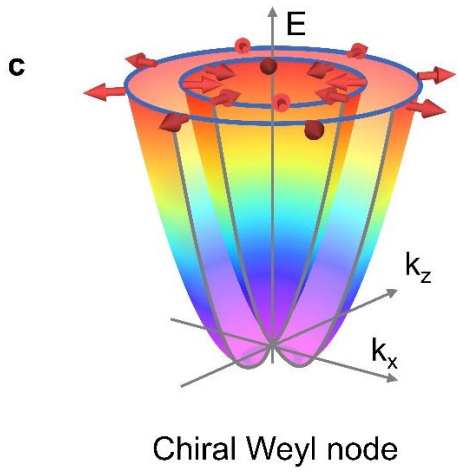
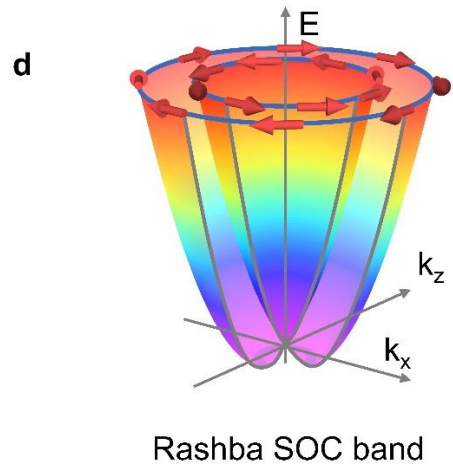

Chiral Weyl node

Rashba SOC band

**Figure 5**

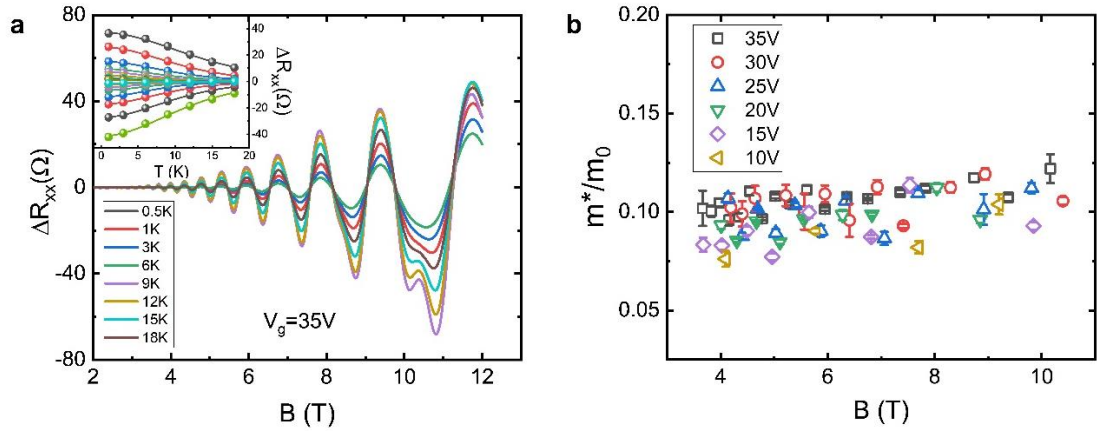

# Supplementary Information for:

# Quantum Hall Effect of Weyl Fermions in Semiconducting *n*-type Tellurene

Gang Qiu[1,2], Chang Niu[1,2], Yixiu Wang[3], Mengwei Si[1,2], Zhuocheng Zhang[1,2], Wenzhuo Wu[3] and Peide D. Ye[1,2]

## List of contents

**Supplementary Note 1**: Additional data from other devices on quantum Hall effects and Shubnikov-de Haas oscillations

**Supplementary Note 2**: Converting Hall resistivity to Hall conductivity with anisotropic Hall tensor

**Supplementary Note 3**: Shubnikov-de Haas oscillations in tilted magnetic field

**Supplementary Note 4**: More data on extracting Berry phase

**Supplementary Note 5**: Weak anti-localization in n-type Te films

---

**Supplementary Fig. 1**: Quantum Hall effect of another high mobility sample.

**Supplementary Fig. 2**: $R_{xx}$ mapping of another high mobility sample in $V_g$-B parameter space.

**Supplementary Fig. 3**: Temperature-dependent SdH oscillations of another high mobility sample.

**Supplementary Fig. 4:** $\rho_{xx}$ and $\Delta\sigma_{xx}$ comparison at various carrier density.

**Supplementary Fig. 5:** Oscillation phase factor extracted from $\Delta\sigma_{xx}$ and $\rho_{xx}$.

**Supplementary Fig. 6:** SdH oscillations in a tilted magnetic field.

**Supplementary Fig. 7:** Extracting Berry phase using SdH oscillation phase offset from 8 more devices.

**Supplementary Fig. 8:** Weak anti-localization effect in n-type Te films.

**Supplementary Note 1:** Additional data from other devices on quantum Hall effects and Shubnikov-de Haas oscillations

To demonstrate the reproducibility of our experimental results, here we present the complete set of data from another high-mobility device. Supplementary Fig. 1 shows the longitudinal ($R_{xx}$) and Hall ($R_{xy}$) resistance measured at 1.7 K up to 31 Tesla. This device also reaches the quantum Hall states at filling factor $\upsilon=4$. The SdH oscillation pattern mapping in Supplementary Fig. 2 shows similarity to the device presented in our main manuscript. Temperature-dependent SdH amplitudes were measured to extract the cyclotron mass (see Supplementary Fig. 3) which is in accordance with our previous results. Additionally, extraction of the oscillation phase offset from 8 more devices is presented in Supplementary Note 4.

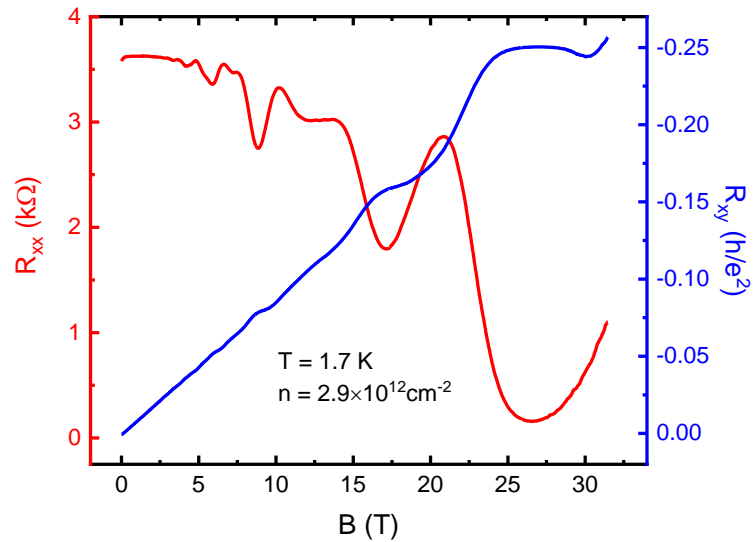

**Supplementary Fig. 1| Quantum Hall effect of another high mobility sample.** Longitudinal ($R_{xx}$, red) and transverse ($R_{xy}$, blue) resistance measured under magnetic field up to 31.4 T at 1.7 K.

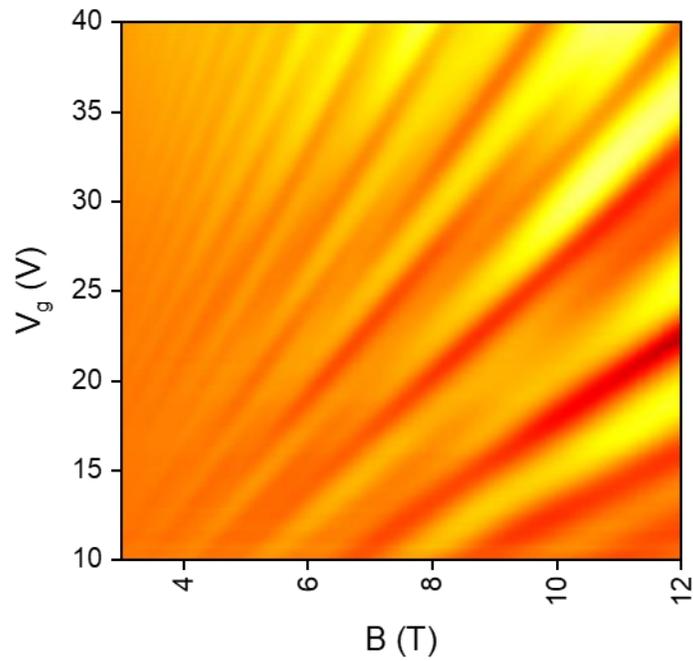

**Supplementary Fig. 2| $R_{xx}$ mapping of another high mobility sample in Vg-B parameter space.**

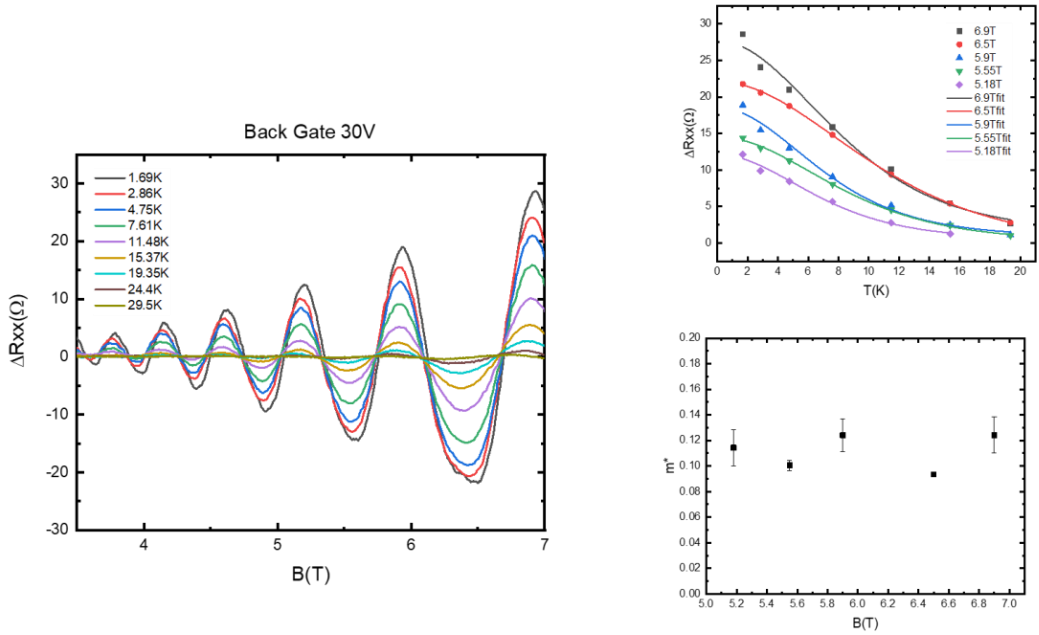

**Supplementary Fig. 3| Temperature-dependent SdH oscillations and extraction of cyclotron mass of another high mobility sample.**

**Supplementary Note 2: Converting Hall resistivity to Hall conductivity with anisotropic Hall tensor**

Although historically it is a common practice to extract the oscillation phase offset from the longitudinal resistivity ($\rho_{xx}$)[1,2], it is argued that the fundamental quantization occurs in the conductivity instead of resistivity[3,4]. Hence using $\rho_{xx}$ instead of $\sigma_{xx}$ can be risky and will sometimes lead to a phase shift[5]. Here we first convert the Hall resistivity into conductivity using the anisotropic Hall tensor and then evaluate the phase offsets extracted using both methods.

Due to its unique atomic structure, the conductivity of Te is highly anisotropic. Extra caution needs to be taken when we apply the Hall tensor to convert from the resistivity matrix to the conductivity matrix. The resistivity ratio $\eta = \rho_{xx}/\rho_{yy} = 1.4$ was previously determined through transport measurements[6,7]. The inverse of the resistivity matrix gives

$$\begin{pmatrix} \sigma_{xx} & \sigma_{xy} \\ \sigma_{yx} & \sigma_{yy} \end{pmatrix} = \begin{pmatrix} \rho_{xx} & \rho_{xy} \\ \rho_{yx} & \frac{1}{\eta}\rho_{xx} \end{pmatrix}^{-1}$$, and subsequently we have $\sigma_{xx} = \frac{\eta\rho_{xx}}{\rho_{xx}^2 + \eta\rho_{xy}^2}, \sigma_{xy} = \frac{\rho_{xy}}{\rho_{xx}^2 + \eta\rho_{xy}^2}$.

Next, we shall evaluate whether using resistivity to construct the Landau fan diagram and extract phase offset is reliable in the case of Te. For some semimetals, because of their high carrier density, the extreme condition $\sigma_{xx} \gg \sigma_{xy}$ gives $\sigma_{xx} \propto \rho_{xx}^{-1}$, hence this will completely flip the phase. However, in our semiconductor system, this extreme condition does not hold. By plotting $\rho_{xx}$ and $\Delta\sigma_{xx}$ (subtracting a smooth background) over a wide gate voltage range (Supplementary Fig. 4), we immediately notice that the minima in both curves coincide, suggesting that the phase shift using $\rho_{xx}$ in this case is negligible and will not undermine our conclusion. As a matter of fact, we notice that using $\Delta\sigma_{xx}$ in the LL fan diagram gives us a slight edge in reliably extracting the phase factor (Supplementary Fig. 5). Hence in the main manuscript, the phase offset is extracted using $\sigma_{xx}$ for correctness, and the rest of the phase offsets included in Supplementary Note 4 are extracted from $\rho_{xx}$ for simplicity.

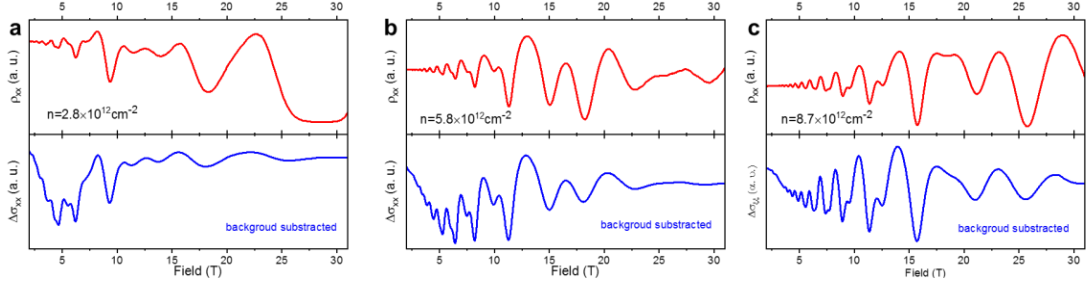

**Supplementary Fig. 4| $\rho_{xx}$ and $\Delta\sigma_{xx}$ comparison at various carrier density.** $\rho_{xx}$ and $\Delta\sigma_{xx}$ (subtracting a smooth background) at carrier density **(a)** $2.8\times10^{12}$ cm$^{-2}$, **(b)** $5.8\times10^{12}$ cm$^{-2}$, and **(c)** $8.7\times10^{12}$ cm$^{-2}$. The minima in $\rho_{xx}$ and $\Delta\sigma_{xx}$ coincide in all three situations.

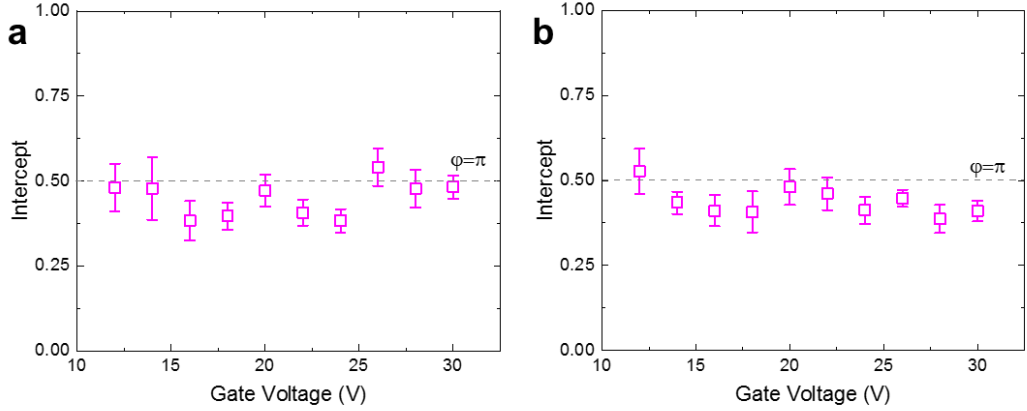

**Supplementary Fig. 5| Oscillation phase factor extracted from (a) $\Delta\sigma_{xx}$ and (b) $\rho_{xx}$.**

## Supplementary Note 3: Shubnikov-de Haas oscillations in tilted magnetic field

To extract the effective g-factor, SdH oscillations are measured under tilted magnetic fields to disentangle the cyclotron energy $E_C = \frac{\hbar eB_\perp}{m^*}$ and the Zeeman energy $E_Z = g^*\mu_B B_{total}$. By rotating the sample to certain angles, one should expect to see the coincidence effect – the exchange of SdH maxima and minima under the condition $E_C = E_Z$. Here we measured our sample in tilted magnetic field, however coincidence effect was not observed until 78.2º (see Fig. S6), from which we can estimate the upper bound of the effective g-factor to be 2.68. The real g* may be even smaller given the fact that no clear trace of developing coincidence effect is observed until such a high angle. This suggests that the Zeeman energy in Te is much smaller than the cyclotron energy, ruling out a possible source of phase shift seen in the case of $WSe_2$[8] or n-type black phosphorous[9].

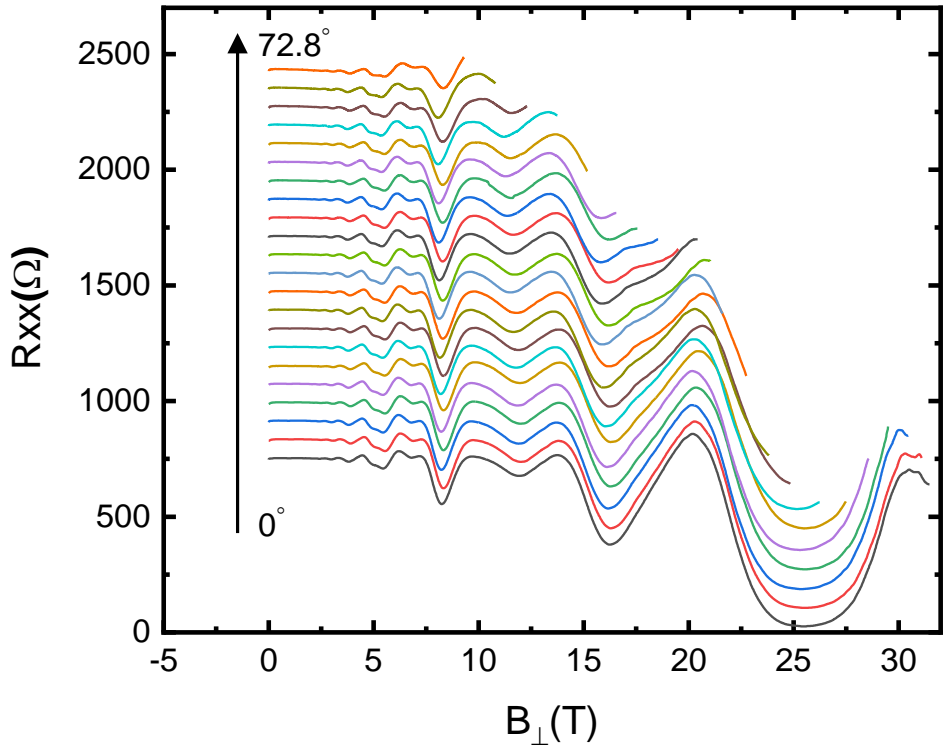

**Supplementary Fig. 6| SdH oscillations in a tilted magnetic field.** The curves are separated by 80 Ω offset for clarity. No evidence of coincidence effect is observed, suggesting a small effective g-factor.

**Supplementary Note 4: More data on extracting Berry phase**

The way we extract Berry phase relies on accurately reading $R_{xx}$ minima at each filling factor. Here we only use the SdH oscillation periods where the degeneracy is not lifted, and all the minima are extracted from the predominant set of oscillations. However, even the splitting is not resolvable at low magnetic fields, and there is a chance that the energy splitting within the Landau levels can deviate $R_{xx}$ minima and cause misinterpretations when extracting the Berry phase. To eliminate this uncertainty, we extracted Landau fan diagram intercepts for another 8 devices as shown in Fig. S7, and all of them unanimously show a π phase shift. This is solid proof that the phenomenon reported in the main paper is reproducible, and our method of extracting Berry phase is reliable.

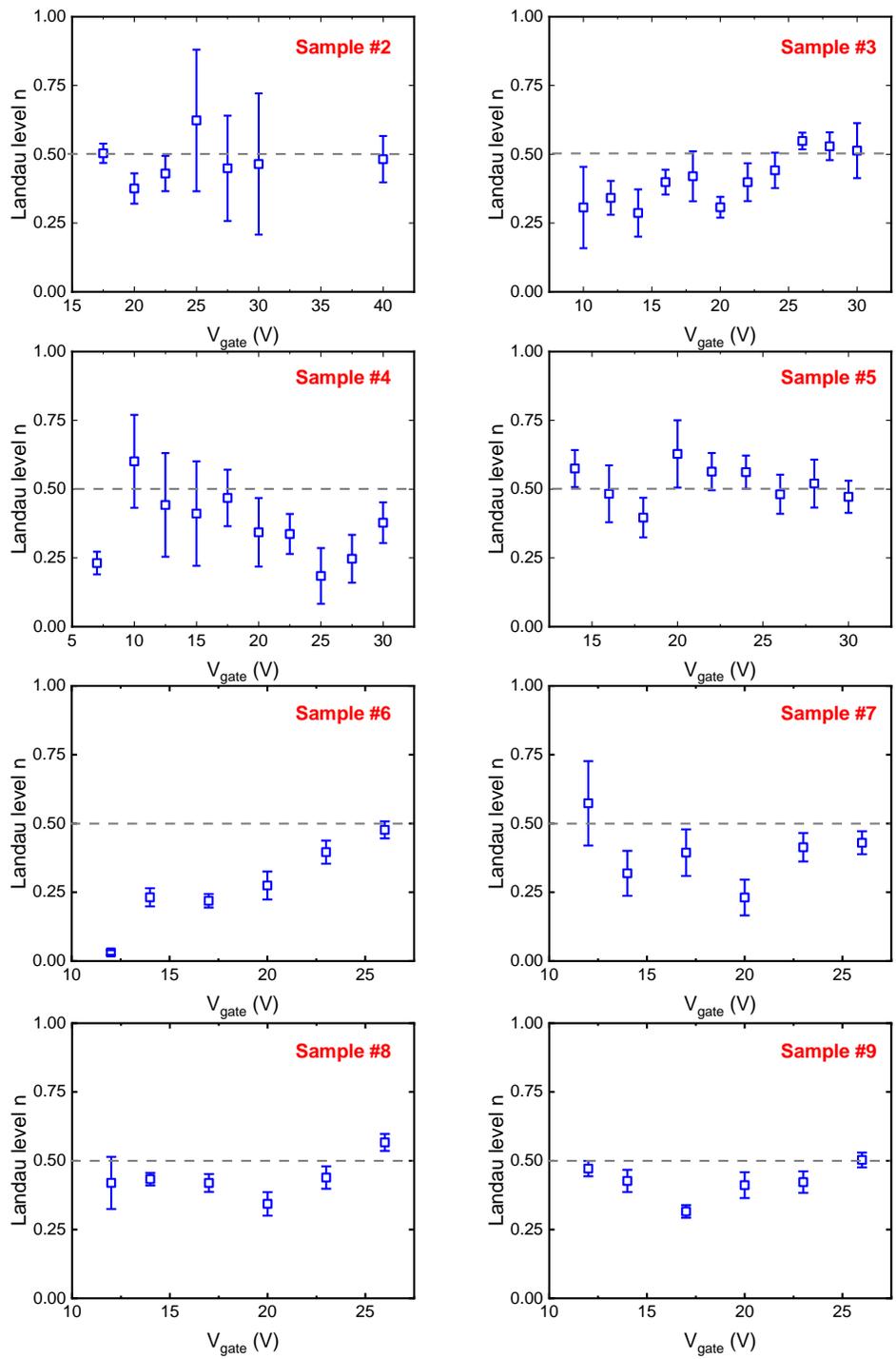

**Supplementary Fig. 7| Extracting Berry phase using SdH oscillation phase offset from 8 more devices.**

**<u>Supplementary Note 5</u>: Weak anti-localization in n-type Te films**

Weak anti-localization (WAL) is a phenomenon in the low B field regime that causes the resistance of a material to show a dip centered at zero. In normal materials there should be a peak in resistance at zero field, since the magnetic field destroys the localization effect and reduces the resistivity of the material, which is referred to as the weak localization effect. However, in systems with strong spin-orbit coupling, the trajectory of the electrons travelling around the disorder clockwise and counterclockwise will contribute a negative sign to the wavefunction and the constructive interference will reduce the resistivity. WAL in p-type Te films has been reported in previous work[10,11]. Here we show some preliminary WAL data in n-type Te films in Fig. S8 to verify strong SOC in the Te system. More comprehensive analysis of n-type Te WAL effects is beyond the scope of this work and will be reported independently[12].

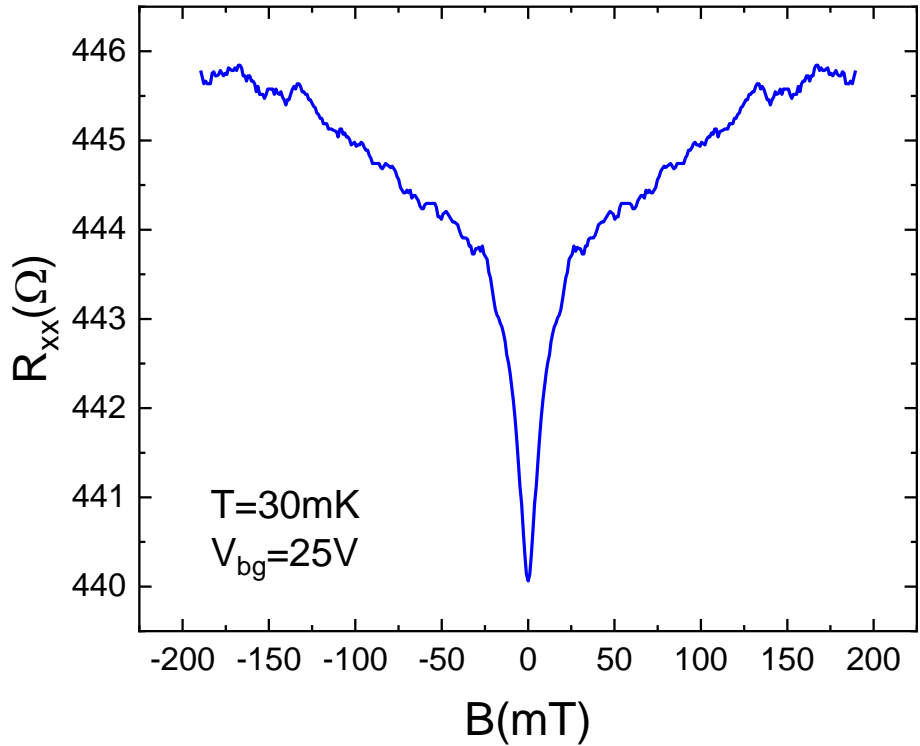

**Supplementary Fig. 8| The weak anti-localization effect in n-type Te films.**

1. Zhang, Y. B., Tan, Y. W., Stormer, H. L. & Kim, P. Experimental observation of the quantum Hall effect and Berry's phase in graphene. *Nature* **438**, 201–204 (2005).

2. Novoselov, K. S. *et al.* Two-dimensional gas of massless Dirac fermions in graphene. *Nature* **438**, 197–200 (2005).

3. Xiong, J. *et al.* Quantum oscillations in a topological insulator $Bi_2Te_2Se$ with large bulk resistivity (6Ωcm). *Phys. E Low-dimensional Syst. Nanostructures* **44**, 917–920 (2012).

4. Ando, Y. Topological Insulator Materials. *J. Phys. Soc. Japan* **82**, 1–32 (2013).

5. Ren, Z., Taskin, A. A., Sasaki, S., Segawa, K. & Ando, Y. Large bulk resistivity

and surface quantum oscillations in the topological insulator $Bi_2Te_2Se$. *Phys. Rev. B* **82**, 241306 (2010).

6. Wang, Y. *et al.* Field-effect transistors made from solution-grown two-dimensional tellurene. *Nat. Electron.* **1**, 228–236 (2018).

7. Qiu, G. *et al.* Quantum transport and band structure evolution under high magnetic field in few-layer tellurene. *Nano Lett.* **18**, 5760–5767 (2018).

8. Xu, S. *et al.* Odd-Integer quantum Hall states and giant spin susceptibility in p-type few-layer $WSe_2$. *Phys. Rev. Lett.* **118**, 067702 (2017).

9. Li, L. *et al.* Quantum oscillations in a two-dimensional electron gas in black phosphorus thin films. *Nat. Nanotechnol.* **10**, 608–613 (2015).

10. Du, Y. *et al.* One-dimensional van der Waals material tellurium: Raman spectroscopy under strain and magneto-transport. *Nano Lett.* **17**, 3965−3973 (2017).

11. Ren, X. *et al.* Gate-tuned insulator-metal transition in electrolyte-gated transistors based on tellurene. *Nano Lett.* **19**, 4738–4744 (2019).

12. Niu, C. *et al.* Gate-tunable Strong Spin-orbit Interaction in Two-dimensional Tellurium Probed by Weak-antilocalization. *arXiv Prepr. arXiv1909.06659* (2019).